\def\eqnum#1{\eqno (#1)}

\documentclass[11pt]{article}
\usepackage{graphics,epsfig,floatflt}

\newcommand{\ka}{\hbox{\ae}}
\begin{document}

\setcounter{page}{49} \sloppy
\noindent {\small\hfill {\bf SPIE Vol. 2798}\\
15th International Conference on Nonlinear Optics {\it ICONO'95} \hfill 27 June - 1
July 1995, St. Petersburg, Russia\\
\centerline { Coherent Phenomena and Amplification without Inversion}} \vspace{12pt}

\begin{center}
{\bf Atomic coherence and interference phenomena in resonant
nonlinear optical interactions}\footnote{Invited review paper}\\
\bigskip
A.K.Popov\\

Institute  for  Physics  Russian Academy of Sciences,
 Krasnoyarsk   University  and Krasnoyarsk Technical  University\\
660036, Krasnoyarsk, Russia. Fax:(3912)438923, E-mail: popov@ksc.krasn.ru\\
\medskip
and S.G.Rautian\\
Institute of Automation and Electrometry Russian Academy of Sciences\\
630090, Novosibirsk, Russia
\end{center}
\section{Abstract} Interference effects in quantum transitions, giving rise to
amplification without inversion, optical transparency and to enhancements in
nonlinear optical frequency conversions are considered. Review of the relevant early
theoretical and experimental results is given. The role of relaxation processes,
spontaneous cascade of polarizations, local field effects, Doppler-broadening, as
well as specific features of the interference in the spectral continuum are discussed.\\
\noindent {\bf Keywords:}\  atomic coherence and interference, resonant
 nonlinear interactions, bound-free transitions,
 amplification without inversion, relaxation-induced processes,
local field effects, inhomogeneous broadening,
frequency-conversion, $VUV$ generation
\section{Introduction}
There has been considerable interest recently in the study of
laser-induced quantum coherence and interference, which leads to
fundamental effects in high resolution
nonlinear spectroscopy, to
amplification of radiation without the requirement of
population inversion ($AWI$) and to  resonantly enhanced
refraction at vanishing (without) absorption ($ERWA$),
to coherent  population  trapping and constructive   contributions
in resonantly    enhanced
nonlinear-optical frequency  conversions  and,  at  the  same  time, to
distractive contributions in absorption of  the  fundamental  and
generated radiations $^{1,2}$. Wide range of applications are expected $^3$.

Resonant  nonlinear optical interference  effects
have  been  subject  of  the  extensive  both   theoretical   and
experimental studies since the discovery of masers and lasers (see for
example $^{2}$ and ref. therein).
In this paper we briefly review some early and recent results of Russian
research groups on this topic.
\section{ Resonant nonlinear optical interference processes}
\subsection {Destructive and constructive interference in classical and
quantum optical physics}
Interference is one of the fundamental physical phenomena. Two oscillations
at one and the same, or close, frequencies may interfere both
in constructive and destructive ways. One can manipulate by the
 resulting oscillations  with variation of the relative phase
 and the amplitudes of the interfering oscillators in order to
enhance or, on the contrary, to eliminate the oscillations of any nature.
Interference is widely used in optical physics, including quantum optics.
 The concept of interference is more general, then the notions of
elementary quantum-optical processes, such as one-photon, multistep and
multiphoton transitions. These notions were introduced and classify
at their frequency-correlation properties in the framework
of the perturbation  theory. Indeed, in resonant interactions, these
 properties may be drastically  changed with growth of the intensity
of the coupled fields $^{4;2b,c}$. The latter may give rise to such effects
in nonlinear spectroscopy of Doppler broadened transition, as
compensation of the residual inhomogeneous broadening in Raman-like
and cascade configurations $^{4;2b,c;5}$.

Quantum interference may occur when coherent superposition of real states is involved
in a process$^6$. Alternatively, interfering frequency-degenerate intraatomic
oscillations may originate from different correlated quantum pathways, contributing
in one and the same frequency. For example, in the weak-field approximation,
 these can be one- and two-photon
contributions to an optical process, associated with the radiation
at a given frequency. Such process may be thought as that started
from the coherent superposition of closely spaced real energy-%
level and quasi-level (virtual state), created by the auxiliary
strong field $^{2b,c;4}$. Such a coherent superposition can be produced
even more easily than in the case of real doublet state. In general,
even in the cases, when many elementary processes contribute to an optical
process  and their classification is troublesome, one can explain and
predict experimental results with the aid of the notion of interfering
frequency-degenerated components of nonlinear polarization. The
amplitudes of the components can be varied with the intensities and
phases -- with the frequency-detunings of the driving fields.

\subsection {Effect of energy levels population and relaxation, density
matrix approach }

In general case of open energy-level configuration with all the
levels being populated and various relaxation processes involved,
density-matrix method is the most convenient for the analysis of a
resonant nonlinear-optical response. Explicit formulae, describing spectral
properties of a weak probe field in the presence of an auxiliary strong one,
in cascade, $V$ and $\Lambda$ configurations can be easily
derived in the similar way $^{2b,c}$. We shall show that on the example
of the energy-level schematic, given on Fig.1.

\begin{floatingfigure}{50mm}
\includegraphics[width=45mm]{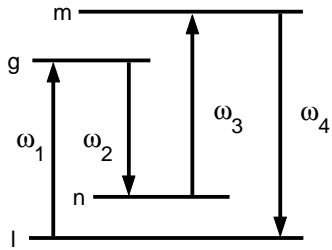}
\caption{Atomic energy levels configuration.}
\end{floatingfigure}
  Fields $E_{1}$ at frequency
$\omega_{1}\approx \omega_{gl}$ and $E_{3}$ at frequency
$\omega_{3}\approx \omega_{mn}$ are strong. Fields $E_{2}$ at frequency
$\omega_{2}$ and $E_{4}$ at frequency
$\omega_{4}$ are probe ones. We shall derive the conditions to achieve $ AWI$
 at the transition $gn$, as well as at transition $ml$, so that both $V$
 and $\Lambda$ configurations are embedded. Frequency of the probe field
may be both higher and lower compared to the driving field.

Consider energy-level configuration, shown in the Fig.1.
 Density matrix equations in the interaction representation, relevant to the
problem under consideration, can be written in the form:

\noindent
$\rho_{lg}=r_1\cdot exp(i\Omega_1t)$,\quad
$\rho_{nm}=r_3\cdot exp(i\Omega_3t)$,\quad
$\rho_{ng}=r_2\cdot exp(i\Omega_1t)+\tilde{r}_2\cdot exp[i(\Omega_1+\Omega_3-\Omega_4)t]$,

\noindent
$\rho_{lm}=r_4\cdot exp(i\Omega_4t)+\tilde{r}_4\cdot exp[i(\Omega_1-\Omega_2+\Omega_3)t]$,\quad
$\rho_{ln}=r_{12}\cdot exp[i(\Omega_1-\Omega_2)t]+r_{43}\cdot exp[i(\Omega_4-\Omega_3)t]$,

\noindent
$\rho_{ii}=r_i$,

\noindent
$P_2r_2=iG_2\Delta r_2-iG_3r_{32}^*+ir_{12}^*G_1$, \qquad
$d_2\tilde{r}_2=-iG_3r^*_{41}+ir^*_{43}G_1$,

\noindent
$P_4r_4=i\left[G_4\Delta r_4-G_1r_{41}+r_{43}G_3\right]$,\qquad
$d_4\tilde{r}_4=-iG_1r_{32}+ir_{12}G_3$\,

\par\noindent
$P_{41}r_{41}=-iG_1^* r_4+ir_1^*G_4$,\qquad
$P_{43}r_{43}=-iG_4 r_3^*+ir_4G_3^*$,

 \noindent
$P_{32}r_{32}=-iG^*_2 r_3+ir_2^*G_3$,\qquad
$P_{12}r_{12}=-iG_1 r_2^*+ir_1G_2^*$,

 \noindent
$\Gamma_m r_m=-2Re\{iG_3^*r_3\}+q_m $,\qquad
$\Gamma_n r_n=-2Re\{iG_3^*r_3\}+\gamma_{gn}r_g+\gamma_{mn}r_m+q_n $,

 \noindent
$\Gamma_g r_g=-2Re\{iG_1^*r_1\}+q_g $,\qquad
$\Gamma_l r_l=-2Re\{iG_1^*r_1\}+\gamma_{gl}r_g+\gamma_{ml}r_m+q_l $,

 \noindent
$\Delta r_1=r_l-r_g,\qquad \Delta r_2=r_n-r_g,\qquad
\Delta r_3=r_n-r_m,\qquad \Delta r_4=r_l-r_m$.

Where
 \noindent
$\Omega_1=\omega_1-\omega_{lg}$,\qquad
$\Omega_3=\omega_3-\omega_{mn}$,\qquad
$\Omega_2=\omega_2-\omega_{gn}$,\qquad
$\Omega_4=\omega_4-\omega_{ml}$,

 \noindent
$G_1=-{\bf E_1}{\bf d}_{lg}/2\hbar$,\quad
$G_2=-{\bf E_2}{\bf d}_{gn}/2\hbar$,\quad
$G_3=-{\bf E_3}{\bf d}_{nm}/2\hbar$,\quad
$G_4=-{\bf E_4}{\bf d}_{ml}/2\hbar$,

 \noindent
$P_1=\Gamma_{lg}+i\Omega_1$,\quad
$P_2=\Gamma_{ng}+i\Omega_2$,\quad
$P_3=\Gamma_{nm}+i\Omega_3$,\quad
$P_4=\Gamma_{lm}+i\Omega_4$,,\quad
$P_{12}=\Gamma_{ln}+i(\Omega_1-\Omega_2)$,\quad
$P_{43}=\Gamma_{ln}+i(\Omega_4-\Omega_3)$,\quad
$P_{32}=\Gamma_{gm}+i(\Omega_3-\Omega_2)$,\quad
$P_{41}=\Gamma_{gm}+i(\Omega_4-\Omega_1)$,

 \noindent
$d_2=\Gamma_{ng}+i(\Omega_1+\Omega_3-\Omega_4)$,\quad
$d_4=\Gamma_{lm}+i(\Omega_1-\Omega_2+\Omega_3)$.

Here $\Omega_i$ are frequency detuning from the resonances,
$G_i$ --- Rabi frequencies, $\Delta r_i$ --- power--depending
population differences, $\Gamma_{ij}$ --- homogeneous half linewidths,
$\Gamma_i^{-1}$ --- lifetimes, $\gamma_{ij}$ --- relaxation rates
from $i$ to $j$ states, $q_i$ --- population rate by a incoherent source.
Density matrix amplitudes $r_i$ determine
absorption/gain and refraction indexes, $\tilde{r}_i$ --- determine
four -- wave mixing driving nonlinear polarizations.

The equations and their solution for the cascade atomic configurations
can be derived by the simple change of the detunings signs $^{2b}$.

\subsection{Laser--induced atomic coherence and classification
 of resonant nonlinear  effects }
Solution of the coupled density -- matrix equations may be represented
 in the form:
 $$ r_{1,3}=i {G_{1,3}\Delta r_1}/{P_1}, \ \ r_{2,4}=i {G_{2,4}}R_{2,4}/{P_{2,4}},
$$ $$R_2=\frac { \Delta r_2(1+g_7+v_7)-v_3(1+v_7-g_8)\Delta r_3-g_3(1+g_7-v_8)\Delta
r_1} {(1+g_2+v_2)+[g_7+g_2(g_7-v_8)+v_7+v_2(v_7-g_8)]},\eqno(1)$$ $$R_4=\frac { \Delta
r_4(1+v_5+g_5)-g_1(1+g_5-v_6)\Delta r_1-v_1(1+v_5-g_6)\Delta r_3}
{(1+g_4+v_4)+[v_5+v_4(v_5-g_6)+g_5+g_4(g_5-v_6)]}, \eqno(2)$$ $$\Delta
r_1=\frac{(1+\ka_3)\Delta n_1+b_1\ka_3 \Delta n_3}
{(1+\ka_1)(1+\ka_3)-a_1\ka_1b_1\ka_3},\ \Delta r_3=\frac{(1+\ka_1)\Delta n_3+a_1\ka_1
\Delta n_1}{(1+\ka_1)(1+\ka_3)-a_1\ka_1b_1\ka_3},$$ $$\Delta r_2=\Delta
n_2-b_2\ka_3\Delta r_3-a_2\ka_1\Delta r_1,\ \Delta r_4=\Delta n_4-a_3\ka_1\Delta
r_1-b_3\ka_3\Delta r_3;$$ $$r_m=n_m+(1-b_2)\ka_3\Delta r_3,\; r_g=n_g
+(1-a_3)\ka_1\Delta r_1,\; r_n=n_n-b_2\ka_3\Delta r_3+a_1\ka_1\Delta r_1,\eqno(3)\;$$
$$r_l=n_l-b_1\ka_3\Delta r_3+a_3\ka_1\Delta r_1,\ \Delta r_i(E_1=0, E_3=0)= \Delta
n_i;$$ $$ g_1=\frac{|G_1|^2}{P_{41}P_1^*}, g_2=\frac{|G_1|^2}{P_{12}^*P_2},
g_3=\frac{|G_1|^2}{P_{12}^*P_1^*}, g_4=\frac{|G_1|^2}{P_{41}P_4},
g_5=\frac{|G_1|^2}{P_{43}d_2^*}, g_6=\frac{|G_1|^2}{P_{41}d_2^*},
g_7=\frac{|G_1|^2}{P_{32}^*d_4^*}, g_8=\frac{|G_1|^2}{P_{12}^*d_4^*},$$
$$v_1=\frac{|G_3|^2}{P_{43}P_3^*}, v_2=\frac{|G_3|^2}{P_{32}^*P_2},
v_3=\frac{|G_3|^2}{P_{32}^*P_3^*}, v_4=\frac{|G_3|^2}{P_{43}P_4},
v_5=\frac{|G_3|^2}{P_{41}d_2^*}, v_6=\frac{|G_3|^2}{P_{43}d_2^*},
v_7=\frac{|G_3|^2}{P_{12}^*d_4^*}, v_8=\frac{|G_3|^2}{P_{32}^*d_4^*}; $$
$$\ka_1=\ka_1^0\frac{\Gamma_{lg}^2}{|P_1|^2},
\ka_1^0=\frac{2(\Gamma_l+\Gamma_g-\gamma_{gl})} {\Gamma_l\Gamma_g\Gamma_{lg}}|G_1|^2,
\ka_3=\ka_3^0\frac{\Gamma_{mn}^2}{|P_3|^2},
\ka_3^0=\frac{2(\Gamma_m+\Gamma_n-\gamma_{mn})}
{\Gamma_m\Gamma_n\Gamma_{mn}}|G_3|^2;$$
$$a_1=\frac{\gamma_{gn}a_2}{\Gamma_n-\gamma_{gn}}=
\frac{\gamma_{gn}\Gamma_la_3}{\Gamma_n(\Gamma_g-\gamma_{gl})}=
\frac{\gamma_{gn}\Gamma_l}{\Gamma_n(\Gamma_l+\Gamma_g-\gamma_{gl})}, $$
$$b_1=\frac{\gamma_{ml}\Gamma_nb_2}{\Gamma_l(\Gamma_m-\gamma_{mn})}
=\frac{\gamma_{ml}b_3}{\Gamma_l(\Gamma_l-\gamma_{ml})}
=\frac{\gamma_{ml}\Gamma_n}{\Gamma_l(\Gamma_m+\Gamma_n-\gamma_{mn})}.$$ By
substituting frequency deviations $\Omega_{i}$ for that Doppler-shifted
$\Omega_{i}-k_{i}v$ ($v$ is atomic velocity) we can take into account the effect of
atomic motion. Imaginary part of density-matrix amplitudes $r_{2}$ and $r_{4}$
represent absorption or gain at the corresponding probe-field frequencies.
 At $G_3=0$ equations (2) and (3) convert in
solutions for $\Lambda$ and $V$ schemes $$r_{2}=i \frac{G_{2}}{P_2}\cdot \frac{\Delta
r_2-g_3\Delta r_1}{1+g_2},\ r_{4}=i\frac{G_{4}}{P_4}\cdot \frac{\Delta r_4-g_1\Delta
r_1}{1+g_4}.\eqno(4)$$

Following$^{2b,c}$ we can classify resonant nonlinear effects as \quad  1) power
saturation of the populations (eq. (4)); \quad  2) strong-field  induced splitting of
the probe-field resonances, or ac Stark effect (denominators in eqs. (4));   and
\quad 3) nonlinear interference effect ($ NIEF$) (second and third terms in the
nominators of eqs. (2)).
\section{Difference in absorption and emission spectra due to the nonlinear
in\-ter\-fe\-rence effects, amplification without inversion, resonance--enhanced
refraction without absorption} Power of emitted or absorbed radiation, for example at
the frequency $\omega_2$,  which is proportional to $Re(-iG^*_2  r_2)$, can be
considered as a difference between pure emission (the term, proportional to $r_g$)
and pure absorption (the rest terms in eqs.$(2)$ ). The difference in
frequency-dependence of these terms, induced by the auxiliary driving field, is the
origin of $AWI^{2b}$. Refractive index at $ \omega_2$ is determined by
 $Im(-iG^*_2  r_2)$ and, in general, laser-induced minimum in absorption
may coincide with the resonance-enhanced maximum in refraction $ ^{1,3}$.
Thus, {\it laser-induced resonance splitting and $NIEF$
 transform only
spectral shape of absorption/gain  and refractive indices, give rise to
 difference in the line shapes of spontaneous (or pure induced) emission
and absorption, but do not affect
the integral intensity of the spectral lines} $^{2b,c}$:

$$
\int d\Omega_2 Re(-ir_2/G_2)  = \Delta r_2,\qquad
\int d\Omega_4 Re(-ir_4/G_4) = \Delta r_4.
\eqnum{5}$$

\noindent {\it Indeed, $ NIEF$ give rise to electromagnetically induced transparency
 ($EIT$) and to $ AWI$ at the transitions $gn$ (or $ml$),
 when contributions of second and
third terms in the nominators of eqs.$ (2)$ are equal or dominate over
$\Delta r_2$ (or over $\Delta r_4$), correspondingly. From the above presented density-matrix
equations one can see that the coherence at the transitions $gm$ and $ln$
($r_{32}$ and $r_{12}$), induced in cooperation of the strong and the probe
fields, is the source of the $ EIT$ and $ AWI$ effects.\/}

A great number of elementary processes, introduced and defined for the
bare states in the framework of the perturbation theory, may give contribution
to the absorption/gain index $\alpha (\Omega_i)$. Consider, for example,
$\alpha (\Omega_4)$  at the frequency
 $\omega_4>\omega_1$ (Fig.1), reduced by it's maximum value $\alpha^{0}(0)$
 in the absence of the all strong fields, for the case when $E_3=0$. From the
eqs.$(2)$ one finds:
$$
{\alpha (\Omega_4)\over \alpha^{0}(0)}= Re\{ {\Gamma_4\over P_4}\cdot
{\Delta r_4-g_1\Delta r_1\over\Delta n_4( 1+g_4)}\}
\eqnum{6}$$

\noindent
Consider two subcases:

{\it a.Off resonance:}\quad ${|\Omega_1|\approx|\Omega_4|>>\Gamma_1,\Gamma_4;
|g_4|<<1; |g_1|<<1; P_4\approx i\Omega_4; P_1\approx i\Omega_1\approx
i\Omega_4}$.\\
Eq.$(6)$ takes the form:

$$
{\alpha (\Omega_4)\over \alpha^{0}(0)}
\approx
 {\Gamma_4^2\Delta r_4\over\Omega_4^2\Delta n_4} -Re\{{\Gamma_4
(\Delta r_4g_4+\Delta r_1 g_1)
\over i\Omega_4\Delta n_4 }\}\hfill\break
 \approx {\Gamma_4^2\Delta r_4
\over\Omega_4^2\Delta n_4}-{\Gamma_4\Gamma_{14}\over
\Gamma_{14}^2+(\Omega_4-\Omega_1)^2}\cdot{|G_1|^2(\Delta r_1-\Delta r_4)
\over \Omega_4^2\Delta n_4} = \hfill\break
$$
$$
 ={\Gamma_{lm}^2( r_l-r_m)\over (n_l-n_m)\Omega_4^2}-
{ \Gamma_{gm}\Gamma_{lm}\over
\Gamma_{gm}^2+(\Omega_4-\Omega_1)^2}\cdot{|G_1|^2 ( r_m- r_g)
\over \Omega_4^2 (n_l-n_m)}
\eqnum{7}$$

The last terms in eqs.$(7)$ describe Raman-like coupling and originate
both from the nominator and the denominator in eq.$(6)$.
{\it Population inversion between initial
 and final bare states ($r_m=n_m>r_g$) is  required for amplification
 of the probe field}.

{\it b.Resonance}:\quad ${\Omega_1=\Omega_4=0}$.\\
Conditions for $ AWI$ and $ EIT$ are:

$$
g_1\Delta r_1\geq \Delta r_4,\quad or\  {|G_1|^2\over \Gamma_{lg}
\Gamma_{gm}}\cdot (r_l-r_g)\geq
r_l-r_m
\eqnum{8}$$

Eq.$(8)$ shows that {\it due to $ NIEF$, population inversion between initial
 and final bare states is not required in order to attain $ AWI$
 in this case. Small relaxation rate of the coherence, induced in cooperation
of the driving and probe fields, compared to the other relaxation rates
 is the most important}.

Analysis  of the condition for $ EIT$ and $ AWI$ as
 well as of sign--changing line shape in the more details can be found
 in ref.$^{2b}$ both
for open and closed ($l$ is ground state) atomic configurations.
The analysis shows strong dependence of the line shape on the ratios of
both population and coherence relaxation rates as well as on the ratios
of initial unsaturated population differences on the coupled transitions.

\subsection{ Constructive
and destructive interference due to the atomic velocity distribution}

Furthermore,  the analysis shows that the contributions of
 the coherence driving
fields to the spectra may be both constructive and
destructive, depending on the detunings of the probe as well as
of the strong fields. This indicates that in gases with inhomogeneous
 broadening of the coupled transitions, dominating over homogeneous
one, conditions for $AWI$ and $EIT$ may considerably differ from that
for atoms at rest. Nevertheless, it was found out that under certain
conditions sign-changing spectral profiles may be produced too $^{2b,c; 7a,b}$.
At weak intensities of driving field narrow structure, superimposed on
the Doppler background, appears. The shape of the structure is anisotropic
and depends on the angle between the wave vectors of the interacting
radiations. Optically-pumped unidirectional-emitting ring laser may operate
by that. The line shape is also dependent on the intensity of the driving
 field and velocity-changing collisions. Special features may occur,
when some of the coupled transitions are homogeneously, and some of them
are inhomogeneously broadened. It was found out that destructive or constructive
character of the effect of Maxwell's velocity distribution depends on the
fact whether a frequency of the probe field is less or greater
than that of the strong one too.
Analytical results describing general behavior of the
 velocity-averaged functions
 for some limiting cases,including Rabi frequencies larger then homogeneous
linewidths, can be found in ref.$^{2a,b,c; 7a,b}$.

\section{Coherence and nonlinear-optical conversion,
 Enhancements in nonlinear-optical conversion due to multiple resonance and
 induced transparency, Local-field effects }

Nonlinear optical response of a medium experiences a giant
enhancements in one- and multiphoton resonances. This reduces
required fundamental powers down to $cw$ regime $^8$, however imposes
severe limitations on the number density of the medium due to
absorption of fundamental and generated radiations. As it is
discussed above, in the presence of a strong electromagnetic
radiation resonances for a weak probe radiation experience
splitting $^{2,9}$, which exhibits itself in a different ways in real and
imaginary parts of linear and nonlinear susceptibilities. Later
makes possible to {\it combine decrease in absorption with increase in
squared module of nonlinear susceptibilities}, responsible for
optical generation, and at the same time with {\it improvements in
phase-matching and increasing density of the medium} $^{2f,10}$.

With the increase of the atom number density, {\it local field},
acting on an atom, may pretty much differ from the external field
both in the amplitude and phase. This {\it may drastically change shape
of nonlinear spectroscopic structures, including
electromagnetically induced transparency}$^{11,12}$.

 Consider experimental schematics,proposed in ref.$^{13}$, that
 combines the advantages of
both {\it multiple resonance enhancements and increase in atom number density of
nonlinear medium} due to the above mentioned Autler-Townes ($ac$ Stark splitting) as
well as {\it local-field effects}.

\begin{figure}
  \centering
\includegraphics[width=.3\textwidth]{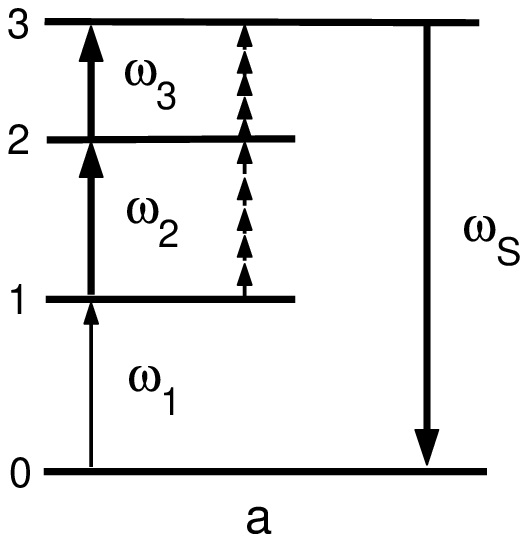}
\includegraphics[width=.3\textwidth]{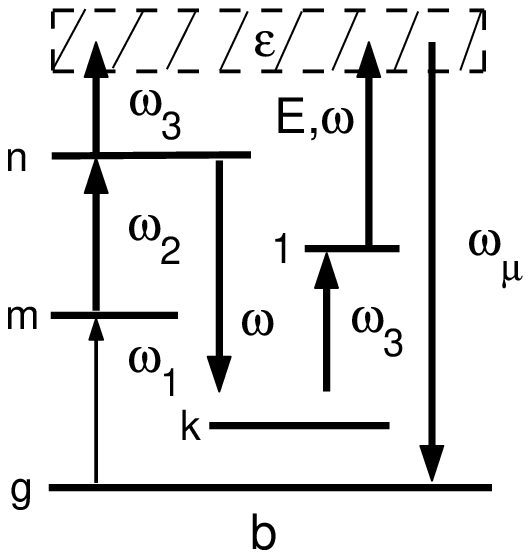}
\includegraphics[width=.3\textwidth]{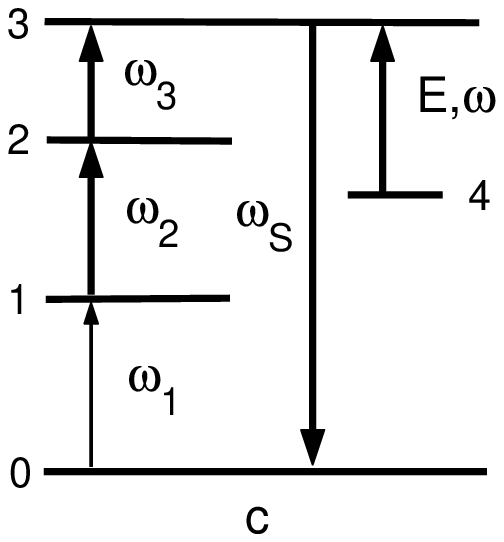}\\
  \caption{Interference enhanced frequency--mixing.\ \ $a.$Triple resonance
 enhanced frequency-conversion. Levels $2$ and $1$  as well as $3$ and $2$
are coupled by the strong either one- or multiphoton interactions. Levels $1$ and $2$
are coupled by the weak field.\ \ $b.$ Enhancement
 in frequency-mixing due to the autoionizing-like resonance, induced in spectral
 continuum by the auxiliary strong fields $E$ and $E_3$.\ \  $c.$ Enhancement in
 three-photon resonant four-wave mixing, induced by the auxilliary strong field $E$.}
\end{figure}

Consider energy-level scheme, shown in Fig.$2a.$ Strong
 fields at frequencies $\omega _{3}$ and $\omega _{2}$ couple unpopulated
 levels $3$ and $2$
(Rabi frequency $G_{3})$ and 2 and 1 (Rabi frequency $G_{2})$, respectively.
Field at $\omega _{1} \approx  \omega _{10}$ as well as generated
$\omega _{s} = \omega _{1}+ \omega _{2}+ \omega _{3}$ are weak, do
not change populations of the levels and are accounted for only in
the lowest order of the perturbation theory. Absorption and
refraction indexes for the probe fields at $\omega _{1}$
and $\omega _{s}$ are represented
 by the imaginary and real parts of
$$
\chi _{1}(-\omega _{1};\omega _{1}) = (\chi ^{0}_{1}/ P_{01}) f_{1},
\qquad \chi _{s}(-\omega _{s};\omega _{s}) = (\chi ^{0}_{s}/ P_{03}) f_{s}
\eqnum{9}$$

Nonlinear susceptibility is:
 $$\chi ^{NL} (-\omega _{s};\omega _{1}+ \omega _{2}+ \omega _{3})
= (\chi ^{NL}_{0}/P_{01} P_{02} D_{03}) f,
\eqnum{10}$$
where $\chi ^{0}_{1},  \chi ^{0}_{s}$ and $\chi ^{NL}_{0}$
 are resonant values of the susceptibilities at negligibly
small $G_{2}$ and $ G_{3}$. Factors $f_{1}$, $f_{2}$ and $f$
 describe effects of the strong fields.  Simple density - matrix calculations,
 similar to given in$^{2b,e;10a.}$  yield:
$$
f_{1}= \{1 + g_{2}/ P_{01} P_{02} [1+ (g_{3}/ P_{02}D_{03})]\}^{-1},
\eqnum{11}$$
$$
f_{s}= \{1 + g_{3}/ P_{03} D_{02} [1+ (g_{2}/D_{02}D_{01})]\}^{-1},
\eqnum{12}$$
$$
f =  f_{1} [1 + g_{3}/  D_{03} P_{02}]^{-1} = [1+ (g_{2}/D_{02}D_{01})
+ (g_{3}/  D_{03} P_{02})]^{-1}
\eqnum{13}$$
$$
P_{01}= 1 + ix_{1},\quad P_{02} = 1 + ix_{0},\quad P_{03}= 1 + ix_{s};
\quad D_{01}= 1 + iy_{1},\quad D_{02} = 1 + iy_{0},\quad D_{03}= 1 + iy_{s};$$
$$ x_{1}= (\omega _{1}- \omega _{10})/\Gamma _{10} = 0,
\quad x_{02}= (\omega _{1}+ \omega _{2}- \omega _{21})/\Gamma _{20} = 0,
\quad x_{s}= (\omega _{s}-\omega _{30})/\Gamma _{30} =0;$$
$$ y_{1}= (\omega _{s} - \omega _{3} - \omega _{2}-
 \omega _{10})/\Gamma _{10} = 0,
 \quad y_{02}= (\omega _{s} - \omega _{3}- \omega _{21})/\Gamma _{20} = 0,
 \quad y_{s}= (\omega _{1} + \omega _{2}
  + \omega _{3} - \omega _{30})/\Gamma _{30} =0;$$
$$ g_{2}= G^{2}_{2} /\Gamma _{10}\Gamma _{2},
\quad g_{3}= G^{2}_{3} /\Gamma _{30}\Gamma _{20},
$$
$ \Gamma _{ij}$ are homogeneous halfwidth of the corresponding
 transitions. In the case, when $E_s$ is not a probe field, but generated
radiation, $\omega_s = \omega _{1} + \omega _{2}  + \omega _{3}$  and
$D_{0i} = P_{0i}$.

Factors $f_{1}, f_{s}$ and $f$ are different and describe splitting of the corresponding
resonances. Frequency-dependence and difference from unity of the factors
 $f_{1}, f_{s}$ and $f$ is determined by the coherence, induced at the transition
$02$ by the two combinations of strong and weak fields $(E_{1}, E_{2}$ and $E_{s},
E_{3}$). Generated power $P \propto  g_{2}g_{3}\mid \chi ^{NL}\mid ^{2}$,
 depends not only on imaginary
but on real part of $\chi ^{NL}$ too, and because of that may not
 deplete in the spectral range of induced transparency
and phase-matching.
 Each resonance increases $\mid \chi ^{NL}\mid ^{2}$ by the factor
 of $x^{-2}_{i}$ , which
may be on the order of $10^{6}$. Laser-induce spectral structures in
real parts of $\chi _{1}$ and $\chi _{s}\/ ($dispersion caused by the coherence
at the $02$ transition), provide additional means to phase-match frequency -
conversion by the small detunings of the fundamental radiations
from the resonances. Triple resonance may yield total enhancement in
generated power on the
order of $10^{18}$. Due to the induced transparency, number density of
the atoms $N$ and consequently $P \propto  N^{2}$ may be increased
 by  several orders of the magnitude in addition.

At high number density of the atoms, local fields may
significantly differ from the external electromagnetic fields both
in amplitudes and in phases. As it was shown in $^{12,13}$, that may
drastically change spectral properties of the induced transparency
as well as of the generating nonlinear polarization. Similar to $^{11,12}$,
making  use Lorentz-Lorenz approximation, local field effects can
be accounted for by the substituting one- photon resonances on that
red-shifted (by substituting $x_{1}$ and $x_{s}$ for $x_{1}+ C_{1}$ and
 $x_{s} + C_{s},\quad C_{1}= N\mid d_{10}\mid ^{2}/3\epsilon _{0}
 \Gamma _{10};
 C_{s}= N\mid d_{30}\mid ^{2}/3\epsilon _{0} \Gamma _{30}$,
 $\epsilon _{0}$- is permittivity of free
space) . Due to the fact that this does not influence transition
frequencies between the excited states and that of the multiphoton
transitions, the introduced shifts may drastically change effects
of strong electromagnetic fields at $\omega _{2}$ and $\omega _{3}$
on both dressed linear and nonlinear responses.

Equations, given above, can be easily generalized on the cases
of the {\it higher order processes}. For example, when $1$--$0$
and/or $3$--$2$,
$ 2$--$1$ are multiphoton transitions, generalization can be done
simply by substituting one-photon Rabi frequencies and detunings for the corresponding
multiphoton ones. Manipulations by the nonlinear susceptibility, absorption
and refractive indexes for the generating radiation with the
{\it auxilliary} strong fields,  coupled to the {\it adjacent}
 transitions (both bound and continuum states, Figs. $2 b,c.$),
 were considered in ref.$^{2f,10}$.

    \section{Nonlinear interference effects at bound-free
 transitions, Laser-induced autoionizing-like resonances (laser induced continuum
 structure)}

Nonlinear interference phenomena, similar to those at bound-bound
transitions, including $AWI$ and $EIT$,  can occur at the transitions
 to ionization continuum. Appropriate theory was developed in ref. $^{2f,10a,14}$.
 Similar case, relevant to the zone bands in crystals, was considered
in ref.$^{15}$. Laser induced autoionizing like resonances -- laser induced
continuum structure ($LICS$) was observed in the experiments ref.$^{16}$, and
since the end of $80's$ studies of the resonant interference processes
 in the context of $LICS$, $AWI$ and $EIT$, first at bound--free and then
at bound--bound transitions, have involved a number of research
 groups$^{17,18}$.

Potential feasibilities to manipulate both by $LICS$ and by
the splitting of the
discrete resonances in order to enhance short - wavelengths
 frequency - mixing output  and to decrease resonant  absorption of the both
 fundamental and generated radiations can be shown with the example of
Fig.$2b.$, generalized for the case, when $\omega_1$ is close to
 $\omega_{10}$, and radiations at $\omega_2$, $\omega_3$  and $\omega$
are strong. The example combines opportunities to manipulate by two $LICS$
and by depletion of absorption at the discrete transitions. Contribution of
strong off - resonant $k$ levels are taken into account too.
By that, the detunings $\mid \omega _{1}$-$\omega _{gm}\mid $,
$\mid \omega _{1}+\omega _{2}-\omega _{gn}\mid $ and
$\mid \omega -\omega _{3}-\omega _{nl}\mid $
are assumed being much less than all the rest.
Density - matrix calculations give the expressions for  nonlinear
 susceptibility
$\chi ^{(3)}(\omega _{\mu }=\omega _{1}+\omega _{2}+\omega _{3})$, which
determines generated power at the frequency $\omega _{\mu }$, as well as for
 absorption indexes $\alpha (\omega _{1})$ and $\alpha (\omega _{\mu })$
 for probe radiations at corresponding frequences as follows $^{19}$:

$$
\chi ^{(3)}(\omega _{\mu }=\omega _{1}+\omega _{2}+\omega _{3})/\chi ^{(3)}_{0\mu }=K/(D_{gm}X),
\eqnum{14}$$
$$
\alpha (\omega _{1})/\alpha _{01}=Re\{[1-g_{mn}/(D_{gm}X)]/D_{gm}\},
\eqnum{15}$$
$$
\alpha (\omega _{\mu })/\alpha _{0\mu }=1-k_{3}\beta _{l}+k_{3}\beta _{l}(y_{l}+q_{gl})^{2}/(1+{y_{l}}^{2})-
$$
$$
-Re\{k_{4}g_{nn}A^{2}(1-iq_{gn})^{2}/Y\}
\eqnum{16}$$
\noindent where $\chi ^{(3)}_{0\mu}$,
 $\alpha _{01}$ and $\alpha _{0\mu}$ - are
corresponding resonant values at the intencities of all the fields beeing
negligibly weak. The rest parameters are as follows:
$$
K=1-k_{1}\beta _{l}[(1-iq_{nl})(1-iq_{lg })]/[(1-iq_{ng})(1+ix_{l})],
\eqnum{17}$$
$$
A=1-k_{1}\beta _{l}[(1-iq_{ln })(1-iq_{gl})]/[(1-iq_{gn})(1+iy_{l})],
\eqnum{18}$$
$$
X=(1+g_{nn})[1+ix_{n}+g_{mn}/D_{gm}(1+g_{nn})-k_{2}\beta _{l}\beta _{n}(1-iq_{nl})^{2}/(1+ix_{l})],
\eqnum{19}$$
$$
Y=(1+g_{nn})[1+iy_{n}+g_{mn}/p_{gm}(1+g_{nn}-k_{2}\beta _{l}\beta _{n}(1-iq_{nl})^{2}/(1+iy_{l})],
\eqnum{20}$$
$$
D_{gm}=1+i(\omega_1-\omega_{gm})/\Gamma_{gm},\quad p_{gm}=1+i(\omega_\mu-\omega_3-\omega_2-\omega_{gm})/\Gamma_{gm},
\eqnum{21}$$
$$
x _{l}=(\omega_1+\omega_2+\omega_3-\omega-\omega_{gl}-\delta_{ll})/(\Gamma_{gl}+\gamma_{ll}),\quad
x _{n}=(\omega_1+\omega_2-\omega_{gn}-\delta_{nn})/(\Gamma_{gn}+\gamma_{nn}),
\eqnum{22}$$
$$
y _{l}=(\omega_\mu-\omega-\omega_{gl}-\delta_{ll})/(\Gamma_{gl}+\gamma_{ll}),\quad
y_{n}=(\omega_\mu-\omega_3-\omega_{gn}-\delta_{nn})/(\Gamma_{gn}+\gamma_{nn}),
\eqnum{23}$$
$$
k_1=(\gamma_{gl}\gamma_{ln})/(\gamma_{gn}\gamma_{nn}),\/
k_2=(\gamma_{nl}\gamma_{ln})/(\gamma_{ll}\gamma_{nn}),\/
k_3=(\gamma_{gl}\gamma_{lg})/(\gamma_{gg}\gamma_{ll}),\/
k_4=(\gamma_{gn}\gamma_{ng})/(\gamma_{gg}\gamma_{nn}),
\eqnum{24}$$
$$
g_{mn}=\mid{G_{mn}}\mid^2/\Gamma_{gm}\Gamma_{gn},\  \beta _{l}=g_{ll}/(1+g_{ll}),\  \beta _{n}=g_{nn}/(1+g_{nn}),
\eqnum{25}$$
$$
g_{ii}=\gamma_{ii}/\Gamma_{gi},\
q_{ij}=\delta_{ij}/\gamma_{ij},\
\gamma _{ij}=\pi \hbar G_{i\epsilon }G_{\epsilon j}|_{\epsilon =\hbar \omega _{\mu }}+Re\{\sum_k{G_{ik}G_{kj}/p_{gk}}\},
\eqnum{26}$$
$$
\delta _{ij}=\hbar P\int d\epsilon  G_{i\epsilon }G_{\epsilon j}/(\hbar \omega _{\mu }-\epsilon )+Im\{\sum_k{G_{ik}G_{kj}/p_{gk}}\}
\eqnum{27}$$
Factors $0\geq k_i \geq 1$, depending on whether continuum states are
 not degenerate or degenerate (unity).

Comparing eqs.$(14)$ and $(16)$ with corresponding equations from ref.$^{10a, 2f}$,
one can see additional interference $LICS$ structures in generating nonlinear
polarization, absorption and refraction indexes, produced in cooperation by
the $E_3$ and $E$ fields (terms, proportional to $\beta_n$ and $g_n$), which
provide with the supplementary means in absorption spectroscopy and
 for enhancements of generated short-wavelength radiation.

\section{Relaxation-induced coherence processes}

As it was outlined above, relaxation may influence interference processes
both in negative and positive ways. Consider examples, when role of
 relaxation is positive.
 \subsection{$AWI$ due to interference in
 spontaneous cascade of polarizations}

The features in absorption and emission spectra, discussed above, are caused
by interference of contributions of probe field and combination of
 probe and auxiliary strong field in atomic polarization.  As it was outlined
 above, there may be other sources
of interfering intraatomic oscillations. One of the means to obtain $AWI$
 without making use of auxiliary strong fields has been suggested recently
in ref.$^{20}$. The origin is interference through the correlations in
spontaneous decay.

\begin{figure}
  \centering
\includegraphics[width=.25\textwidth]{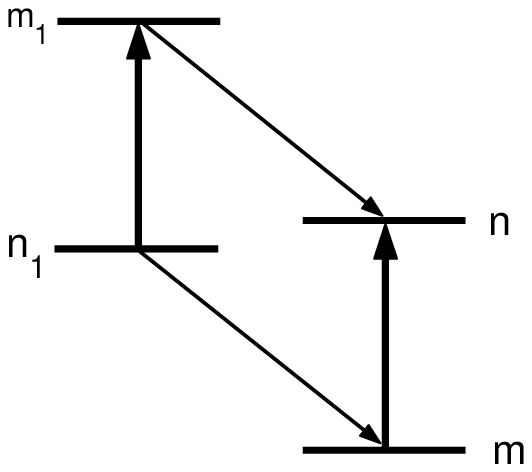}\qquad
\includegraphics[width=.35\textwidth]{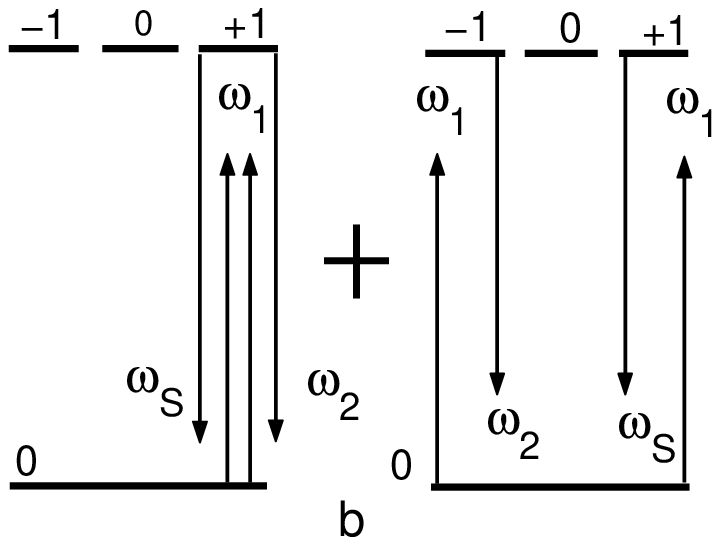}\\
  \caption{Energy-level schematics for relaxation-induced coherent processes.
\  $a.$ $AWI$ through spontaneous cascade of polarizations. \ $b.$ Relaxation-induced $FWM$.}
\end{figure}


Consider four-level atomic configuration shown in Fig.$3a.$
 All four transitions
are allowed. Suppose, that the transition frequency $\omega_{mn}$ is close
to $\omega_{m_1n_1}$, and $\omega_{m_1m}$ is close to $\omega_{n_1n}$,
that is difference $\Delta$

$$
 \Delta = \omega_{m_1n_1} - \omega_{mn} = \omega_{m_1m} - \omega_{n_1n}
   \eqnum{28}$$
  \noindent
is small. In this case interference between considered four radiating channels
is possible. It is caused by the coherence transfer due to interaction with
 the vacuum  oscillations, besides the populations decay
and  spontaneous emissions of photons. For the absorption index in the
 frequency range around $\omega_{mn}$  calculations give:

$$ \alpha ( \Omega )={{ \lambda ^2} \over {4 \pi}} \{ {N_{nm} A_{mn}} {\Gamma \over
{\Gamma ^2 + \Omega ^2}} + {N_{n_1 m_1} A_{m_1 n_1}} [{\Gamma _1 \over
{\Gamma _1 ^2 +(\Omega - \Delta ) ^2}} + {{K C} \over {\Gamma \Gamma _1}}
f( \Omega)] \} , \eqno {(29)}
$$

$$
C = \sqrt {A_{m_{1}m} A_{n_{1}n} A_{mn} / A_{m_{1}}n_{1}} , \qquad
K = {(-1) ^{J_{m} + J_{n_1}}} \sqrt {2{J_m} + 1}\  \sqrt {2J_{{n_1}+1}}
\ \{ \matrix {J_m & J_n & 1\cr
J_{n_1} & J_{m_1} & 1} \} , \eqno {(30)}
$$
$$
f(\Omega)= Re{{ \Gamma \Gamma _1} \over {( \Gamma - i \Omega ) [ {\Gamma _1}
- i ( \Omega - \Delta )]}} = {{ \Gamma {\Gamma _1} [ \Gamma {\Gamma _1} -
\Omega ( \Omega - \Delta )]} \over {( {\Gamma ^2} + {\Omega ^2})
[{\Gamma _1 ^2} + (\Omega - \Delta)^2]}} ,  \eqno {(31)}
$$
$$
N_{nm} = (2J_m + 1) (\rho _n - \rho _m) , \qquad  N_{n_1 m_1} =(2J_{m_1} +1)
(\rho _{n_1} - \rho _{m_1}) .   \eqno {(32)}
$$
  \noindent
Here $\Omega = \omega - \omega_{mn}$, $A_{ij}$ -- Einstein coefficients,
$\Gamma$, $\Gamma_1$ -- are line halfwidths for the interfering transitions,
$J_i$ -- energy level momenta, $N_{ij}$ -- population differences.

The interference term is described by the function $f(\Omega)$,
\  $\int f(\Omega) d\Omega =0$ . Coefficient
$K$ is determined by the moments of four levels under consideration and
may vary in the interval ${-1\leq K \leq 1}$. The case $K\geq 0$ corresponds
to constructive interference (enhancements in the oscillations),
the case $K\leq 0$ --- to destructive interference. The analysis of the
 lineshape eq.$29$ shows it sign-changing behavior. For example,
at $\mid \Omega \mid \gg \Delta$

$$
\alpha ( \Omega )={{ \lambda ^2} \over {4 \pi \Omega ^2}}
\{ N_{nm} A_{mn} \Gamma  + {N_{n_1 m_1} A_{m_1 n_1}} ({\Gamma _1 - K C } )
 \} , \eqno {(33)}
$$
  \noindent
 According to eq.{33}, absorption index may occur negative ($AWI$), if
the requirements

$$
K>O, \qquad (KC/ \Gamma _1 -1)N_{n_1 m_1} A_{m_1 n_1} \Gamma _1 > N_{nm} A_{mn} \Gamma
\eqno{(34)}
$$
  \noindent
are met. When $K\leq 0$, $ \Delta = 0$, the condition

$$
(|K|C/\Gamma -1) N_{n_1 m_1} A_{m_1 n_1} \Gamma > N_{nm} A_{mn} \Gamma_1
\eqno {(35)}
$$
  \noindent
means appearance of $AWI$ in the line center $(\Omega = 0)$.
 Similar phenomena may occur
in the spectral range of the doublet  $\omega_{m_1m}$,  $\omega_{n_1n}$.
Thus, in the considered atomic configuration $AWI$ may be provided by the
correlations in the spontaneous decay without any external action.

    \subsection{ Collision-induced  four-wave mixing}

 Consider example, when collisions and
spontaneous relaxation, as well as  external  magnetic field, break
 destructive interference$^{8a}$ . This remove elimination of for-wave mixing process
and provides with the test, selectively sensitive to the specific modes
of relaxation. The experiment was carried out with $He-Ne$ laser, $\lambda
= 1.52 \mu m$, which is resonant to $2s_2$--$2p_4$ transition of $Ne$.
 The upper level consist of three Zeeman's sublevel ($J_1 = 1$), the lower
 one is
singlet ($J_0 = 0$). Fundamental beam consisted of two linear and orthogonal
polarized components $E_1$ and $E_2$, frequency-shift
 $\Delta = \omega_2 - \omega_1$ being
much less than natural transition linewidth. Intensity of the radiation at
$\omega_1$ was much greater then that at $\omega_2$. Collision and magnetic
field sensitive four-wave mixing output $E_s$
at $\omega_s =2 \omega_1 - \omega_2 = \omega_2 -2 \Delta$ and with
 the same polarization as $E_2$ was detected. Growth of the $FWM$ signal
with the increase of collision rate and strength of magnetic field
 was observed, that can be explained as follows.

Each field and emitting nonlinear polarization $P^{NL}(\omega_s)$
  may be represented as
combination of two circular polarized components $P_+^{NL}(\omega_s)$ and
$P_-^{NL}(\omega_s)$. Formulae for each of these components of
nonlinear polarization consist of two terms. One of them describes $FWM$
of the radiations with one and the same polarizations  in two-level
subsystem, another one -- $FWM$ of the waves with opposite polarizations
in three-level Zeeman's subsystem (Fig.$3b$). In the schematic
 under consideration, it turned out, that the two contributions interfere in
the distractive way and completely eliminate each other, provided by
 the relaxation rates of population and quadruple moment (alignment) in
 the upper level are equal. It is obvious that trapping of the spontaneous
 radiation from the upper level, anisotropic collisions, as well as external
 magnetic field break the counterbalance and, therefore, induce $FWM$ output.
Such dependence was observed in the experiments. External magnetic field turns
the second channel into fully resonant double-$V$ schematics.

\section{ Review of early theory and experiments
 on $NIEF$, $AWI$ and related phenomena }

Coherence phenomena in three-level systems were studied since discovery of
{\it masers}\/. Feasibility to attain $AWI$ in these systems was discussed
in some of publications of that period both for microwave$^{21}$ and optical
transitions$^{22}$. $AWI$ in optical two-level systems was predicted in
ref.$^{23}$ and first was observed in radio-frequency transitions$^{24,2d}$.
In optical range $AWI$ and corresponding features in refractive index were
observed in ref.$^{2e,25}$. Studies of coherence and interference phenomena
in quantum  transitions is growing research area, since they are embedded in
many optical processes of basic and practical importance.

\section{Concluding remarks}

As it was outline, interference is basic and very general phenomenon
 of optical physics, which may play a crucial role in many experimental
 schematics of resonant nonlinear optics. Some of such schematics are shown
in the Fig.$4$.

\begin{figure}
  \centering
\includegraphics[width=.2\textwidth]{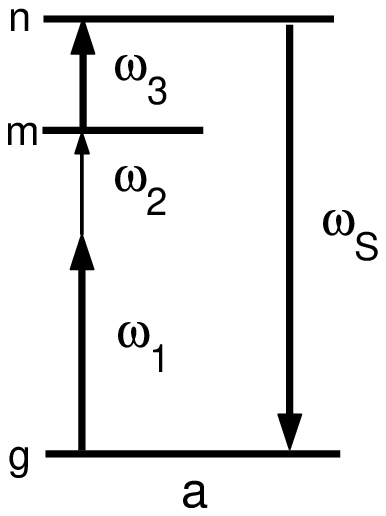}\qquad
\includegraphics[width=.17\textwidth]{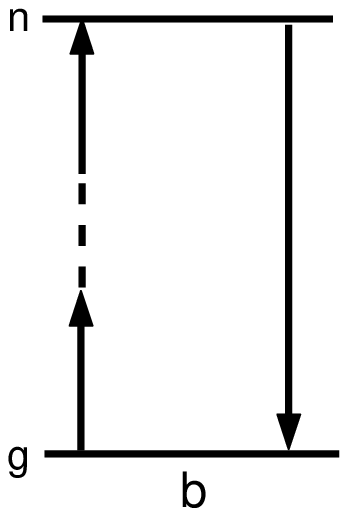}\qquad
\includegraphics[width=.17\textwidth]{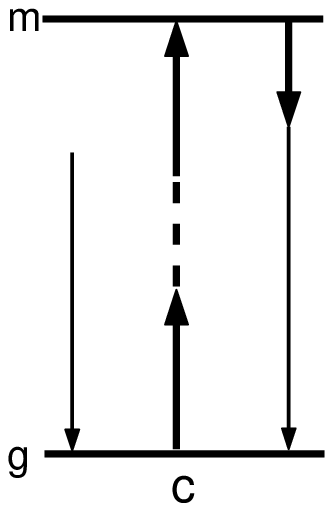}\qquad
\includegraphics[width=.12\textwidth]{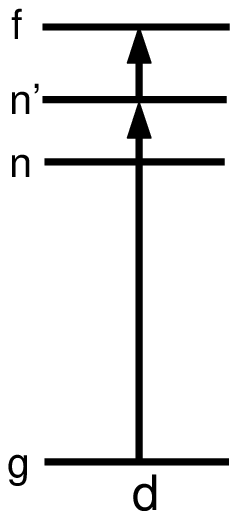}\\
\includegraphics[width=.125\textwidth]{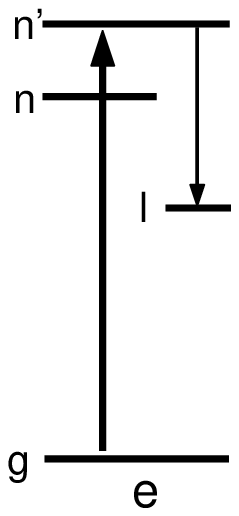}\qquad
\includegraphics[width=.2\textwidth]{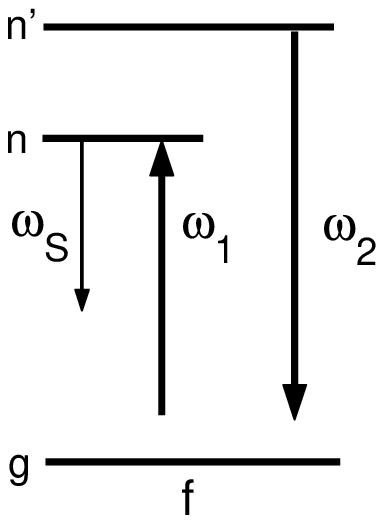}\qquad
\includegraphics[width=.2\textwidth]{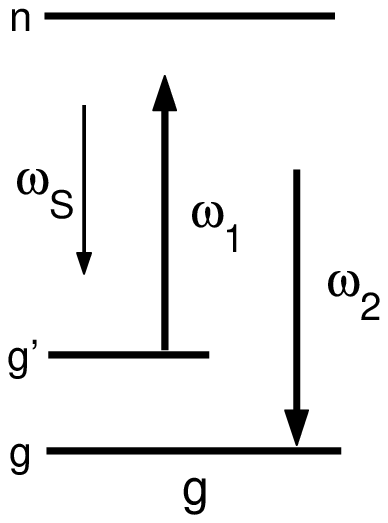}
  \caption{Interfering channels, embedded in resonant  nonlinear optical
processes.}
\end{figure}

 Fig.$4a.$ shows upconversion of weak infrared radiation at the
frequency $\omega_2$.
Fields $E_1$ and $E_3$ are strong. Destructive interference of oscillations
at the frequency $ \omega_{s} -\omega_{3} = \omega_{ng} = \omega_{1} +
 \omega_{2}$ was shown to be one of the main process, limiting the
 conversion efficiency$^{26}$.  Fig.$4b.$ -- interference of multiphoton transition
and one-photon, induced by the generating radiation eliminates population
of the upper level.  Fig.$4c.$ -- off-resonant 7th-order seventh-harmonic
generation interfere with resonant 9th-order seventh-harmonic generation,
that was used for detection of the processes$^{26}$.  Figs.$4d,e.$ -- interference
of contributions of the  doublet sublevels in two-photon and off-resonant
one-photon  transitions.   Figs.$4f,g.$ -- interference of doublet sublevels
in $FWM$.

 Pressure-induced resonance was first proposed and experimentally proved
in$^{8a}$ and later in$^{27}$. {\it The entire analogy} between the schemes {\it 3b}
and {\it 4f} is seen from the formula for the driving coherence (scheme {\it 4f}) $$
\rho^{(2)}_{n'n} \propto V_{n'g}\rho^{(1)}_{gn} + \rho^{(1)}_{n'g}V_{gn} \propto
[{1\over \Omega_2+i\Gamma_{n'g}} - {1\over \Omega_1-i\Gamma_{ng}}] {1\over
\Omega+i\Gamma_{n'n} } = $$ $$ = {1\over(\Omega_2+i\Gamma_{n'g})
(\Omega_1-i\Gamma_{ng})} [1 - i {\Gamma_{nn'} - \Gamma_{n'g} - \Gamma_{ng}\over
\Omega+i\Gamma_{nn'}}]. \eqnum{36}$$ \noindent Here $\Omega_1 = \omega_1 -
\omega_{ng}$, $\Omega_2 = \omega_2 - \omega_{n'g}$, $\Omega = \omega_2 - \omega_1 -
\omega_{n'n}$. At spontaneous relaxation, $\Gamma_{ij} = (\Gamma_i + \Gamma_j)/2$,
and resonance $\Omega = 0$ disappears. Collisions induce this resonance.

\section{Acknowledgements}

This work was supported in part by the International
 Science Foundation
(Grants No 4000, No 4300),  by the  Russian Foundation for
Fundamental Research (Grants N 93-02-03460, N 4300) and
 by the Krasnoyarsk Regional science Foundation, (Grant 4F 0123).

\newpage
\section{References.}

1. For a survey of recent research on AWI and ERWA see:\\
$a.$ M. Fleischhauer, C.H. Keitel, M.O. Scully, Chang Su, B.T. Ulrich,  Shi-Yao Zhu,
Phys. Rev.,
Vol. A46, 1468, 1992; \\
 $b.$ Papers from {\underline {\it "Atomic  Coherence  and  Interference"}} ( Crested
Butte Workshop, 1993), Quantum Optics, Vol. 6, N4, 1994.

2. For the survey of early publications of Russian groups  on  AWI see for
example:\\   $a.$ V.S. Letokhov and  V.P. Chebotaev, \underline {{\it Nonlinear
Spectroscopy}}\/,
Springer- Verlag, 1977;\\
$b.$  A.K. Popov, \underline{{\it Introduction  in  Nonlinear Spectroscopy}} (in
Russ.), Novosibirsk, Nauka, 1983;\\ T.Ya. Popova, A.K. Popov, Zhurn.Prikl.
Spektrosk.,Vol. 12, No 6, 989, 1970\   (Engl.:\ Journ. Appl. Spectr, Vol.12, No 6,
734, 1970), quant-ph/0005047 ;\\ T.Ya. Popova, A.K. Popov, Izv.VUZ, Fizika, No 11, 38,
1970\ (Engl.: Soviet Phys. Journ., Vol. 13, No 11,
 1435, 1970), quant-ph/0005049;\\
T.Ya. Popova, A.K. Popov, S.G. Rautian, R.I. Sokolovskii, Zh. Eksp.
Teor. Fiz., Vol. 57, 850, 1969\  (Engl.: JETP, Vol. 30, 466, 1970,  quant-ph/0005094; \\
$c.$ S.G. Rautian and A.M. Shalagin, \underline{{\it Kinetic  Problems
  of  Nonlinear Spectroscopy}} , North-Holland, 1991;\\
G.E. Notkin, S.G. Rautian, A.A. Feoktistov, Zh. Eksp. Teor. Fiz.,  Vol. 52, No 6,
1673, 1967;\\
$d.$ A.M. Bonch-Bruevich, V.A. Khodovoi, N.A. Chigir, Zh. Eksp. Teor. Fiz.,
 Vol.67, 2069, 1974;\\
$e.$ L.S. Gaida, S.A. Pul'kin, Opt. Spektr., Vol.67, No 2, 761,1989;\\
$f.$ Yu.I. Heller and A.K. Popov, {\underline {\it Laser  Induction  of  Nonlinear
Resonances in Continuous Spectra}\/} , Novosibirsk, Nauka, 1981 (in Russ.)\  (Engl.:
Journ.  Sov. Laser Research, Vol.6, No 1, Jan.-Feb., 1985, c/b Consultants Bureau,
NY, USA).

3. M.O. Scully, Phys.Rev.Lett., Vol. 63, 1855, 1991.

4. T.Ya. Popova, A.K. Popov, S.G. Rautian, A.A. Feoktistov, Zh. Eksp. Teor. Fiz.,
Vol. 57, 444, 1969\  (Engl.: JETP, Vol. 30, 243, 1970), quant-ph/0005081.

5.\   $a.$ C. Cohen-Tannoudji, F. Hoffbeck, S. Reynaud, Opt. Commun., Vol. 27, 71, 1978;\\
 $b.$ A.K. Popov, V.M. Shalaev, Opt. Commun., Vol. 35, 189, 1980.

6.\  $a.$ M.P. Chaika, {\underline {\it Interference of degenerate
atomic states }}\/ (in Russ.), Leningrad University, Leningrad, 1975;\\
$b.$ E.B. Aleksandrov, G.I. Khvostenko, M.P. Chaika,
 {\underline {\it Interference of atomic states }}\/ (in Russ.),
Nauka, Moscow, 1991.

7.\ $a$. M.S. Feld, A. Javan, Phys. Rev., Vol. 177, No 2, 540, 1969;\\
M.S. Feld , In: {\underline {\it Fundamental and Applied Laser Physics}}\/ :
Proceedings of the Esfahan Symposium,\\ August~ 29 --- September 5, 1971/
Ed. by M.S. Feld, A. Javan, N. Kurnit, N.Y., Willey, pp. 369 -- 420, 1973;\\
$b.$ Th. H\" ansch, P. Toschek, Z. Physik, Bd.236, 213, 1970.

8.\  $a.$ Im Tkhekde,{\it et   al}\/, Pis'ma Zh. Eksp. Teor. Fiz., Vol. 24,   8 ,1976;
  Opt.Commun., Vol. 18,  499, 1976; Opt.Commun., Vol. 30, No 2, 196, 1979;
Opt. Spektr.,Vol. 67, No 2, 263, 1989;\\
$b.$ V.M. Klementjev,{\it et   al}\/ , Pis'ma Zh. Eksp. Teor. Fiz., Vol. 24,   8 ,1976;\\
$c.$ L.T. Bolotskikh,  {\it et   al}\/,   Appl.Phys., Vol. B35, 249, 1984;\\
$d.$ V.G. Arkhipkin,{\it et   al}\/, Kvant. Elektron.,Vol. 13, No 7, 1352, 1986.

9.\  $a.$ Yu.M. Kirin,  {\it et   al}\/,
 Pis'ma Zh. Eksp. Teor. Fiz., Vol. 9,  7,  1969;   JETP, Vol. 62,  466, 1972;\\
$b.$ M.P. Bondareva, {\it et   al}\/, Opt. Spektrosk. Vol. 38, 219, 1975.

10.\  $a.$ Yu.I. Heller, A.K. Popov, Opt.Commun., Vol. 18,  449,  1976;  Phys.
Lett., Vol. 50A, 453, 1976;\\
$b.$ S.E. Harris, et al, Phys. Rev. Lett. Vol. 64,1107, 1990;\\
$c.$ V.G. Arkhipkin, Kvant. Elektr., vol. 22, No 1, 81,1995.

11. $a.$ Ch.M. Bowden and J.P. Dowling, Phys. Rev., Vol. A 47, 1247, 1993;\\
$b.$ J.J. Maki {\it et al.}\/,Phys. Rev. Let., Vol. 67, 972, 1993.

12. V.G. Arkhipkin,  A.K. Popov,  A.S. Aleksandrovsky, JETP Lett.,
 Vol.59, No. 6, 398, 1994.

13. A.K. Popov. Triple-resonance enhanced frequency-mixing using electromagnetically
induced transparency. Advance Programme (p.89) and  technical  Digest  of  European
Quantum  Electronics Conference CLEO/Europe-EQUEC 94, 28 August - 2  September,  1994,
Amsterdam, The Netherlands.

14. V.G. Arkhipkin and Yu.I. Heller, Phys. Lett., Vol.98 A, 12, 1993.

15. E.Yu. Perlin, Fiz. Tverd. Tela (Sov. Journ. Sol. State. Phys.), Vol.14, No 7,
2133, 1973; {\it ibid.}\/,Vol. 15, No 1, 66, 1973; Opt. Spektr., Vol.41, No 2, 263,
1976.

16. Yu.I. Heller, V.F. Lukinykh, A.K. Popov, V.V. Slabko, Phys. Lett., Vol. 82 A, No
1, 4, 1981.

17. P.L. Knight, M.A. Lander, B.J. Dalton, Phys. Reports, Vol. 190, No 2, 2, 1990 (and
ref. therein).

18. \ $a.$ S. Cavaliery, {\it et al.}\/, Phys. Rev., Vol. A 47, No 5, 1993; \\
$b.$ O. Faucher, {\it et al.}\/, Phys. Rev. Lett., Vol. 70, No 5, 3004, 1993
 (and ref. therein).

19. A.K. Popov and V.V. Kimberg, (to be published)(Quantum Electron. 28(3), 228-234
(1998).)

20. S.G. Rautian, Pis'ma Zh. Eksp. Teor. Fiz., Vol.61, No 6, 461, 1995.

21. \ $a.$ A. Javan, Phys. Rev.,Vol.107, No 6, 1579, 1957;\\
$b.$ V.M. Kontorovich, A.M. Prokhorov, Zh. Eksp. Teor. Fiz., Vol.33, 1428, 1957;\\
$c.$ T. Yajima, K. Shimoda, J. Phys. Jap., Vol. 15, 1668, 1960;
 Adv.Quant. Electr., p.548, 1961;\\
$d.$ V.M. Fain,{\underline {\it Photons and Nonlinear Media}}\/, Sov. Radio, Moscow,
1972 (p.389), (Vol.1 in: V.M. Fain and Ya.I. Khanin,
 {\underline {\it Quantum Radiophysics}},\/ (In Russ., translated in English).

22. T.Ya. Popova, A.K. Popov, Zh. Eksp. Teor. Fiz., Vol.52, 1517, 1967.

23. S.G. Rautian and I.I. Sobelman, Zh. Eksp. Teor. Fiz., Vol.41, 456, 1961.

24. \ $a.$ D.N. , Yu.S. Konstantinov, V.S. , Izv. Vuz. Radiofiz.,Vol. 8,
513, 1965;\\
$b.$ A.M. Bonch-Bruevich, S.G. Przhibelskii, N.A. Chigir, Vestnik MGU,
 Fisika, Vol.33, No 4, 35, 1978  (Engl.:p. 28).

25. \ $a.$ E.Y. Wu, {\it et al.}\/, Phys. Rev. Lett.,Vol. 38, 1077, 1977;\\
$b.$ I.S. Zelikovich, S.A. Pulkin, L.S. Gaida,
V.N. Komar, Zh. Eksp. Teor. Fiz., Vol. 94, 76, 1988.\\
$c.$ I.M. Beterov, V.P. Chebotaev, Pis'ma v Zh. Eksp. Fiz. Vol. 9, No 4, 216, 1969
(JETP Lett.); explicit statements and description of the observed amplification
without inversion at the transitions of $Ne$ can be found in: I.M. Beterov, Cand.
thesis, Novosibirsk, 1970.

26. \ $a.$ V.G. Arkhipkin and A.K. Popov, {\underline {\it Nonlinear Conversion of
Light in Gases}}\/ (in Russ.), Novosibirsk, Nauka, 1987;\\
$b.$ V.G. Arkhipkin and A.K. Popov,  Sov. Phys. Usp., Vol. 30, No 11, 952, 1987\\
and ref. therein.

27. \ $a.$ Y.H. Zou and N. Blombergen, Phys. Rev., Vol.A34, 2968, 1986;\\
$b.$ L. Rothberg, in {\underline {\it Progress in Optics}}, ed. E.Wolf,
North Holland, Amsterdam, Vol. 24, p.24, 1987\\
and ref. therein.

\end{document}